# Delineation of Raw Plethysmograph using Wavelets for Mobile based Pulse Oximeters

*Sangeeta S. Soni, Yogendra A. Namjoshi*

*Abstract*— The non-invasive pulse-oximeter is a crucial parameter in continuous monitoring systems. It plays a vital role from admission of the patient to surgeries with general anaesthesia. The paper proposes the application of wavelet transform to delineate the raw plethysmograph signals obtained from basic portable and mobile-powered electronic hardware. The paper primarily focuses on finding peaks and baseline from noisy infrared and red waveforms which are responsible for calculating heart-rate and oxygen saturation percentages.

## I. INTRODUCTION

The non invasive plethysmography is widely used method in monitoring the oxygen saturation (SaO2) of the subject. It is one of the quickest and safest tools that helps find heart rate, R-to-R variation and oxygen saturation in the blood. It is one of the important parameter looked-at while admitting the patient and during surgery. There are situations when the oxygen saturation of the patient may decreased beyond a safe threshold and an immediate attention is required. Many multi-parameter monitors stationed near the bed can alarm the medical staff; however these are neither easily portable nor affordable for each bed for majority of rural hospitals. It is also observed that most of the times only few parameters of the multi-parameter monitor are accessed. Thus the pulse-oximeter monitor needs to be portable, simplistic, cost-effective or economical and should be capable of sending alerts to the medical staff. There are widely available compact pulse-oximeters which are finger probe sized. But firstly they have a very small LCD interface for a busy doctor to read. Secondly they work with basic signal processing and electronics and are also battery driven. Segawa et. al. proposes a novel method of using sensor networks for detecting heart rate and heart rate variability [1].

Recently, developments in mobile software applications has enabled features like calling and messaging from within the application which can also connect to the external world based on interfaces like USB, Bluetooth etc. It would be highly efficient to use these techniques to give a missed call for locally available staff when the mobile reads in emergency from the medical device it is talking to. We propose developing such application on mobile that will power the basic electronic hardware connected to the pulse-oximeter probe and read in digitized data (infrared and red signals) to calculate parameters based on processing in software. Furthermore, such can also be embedded inside the probe. It is also proved by A. Tamas *et. al.* that GSM frequency doesn't affect plethysmographic signals[2].

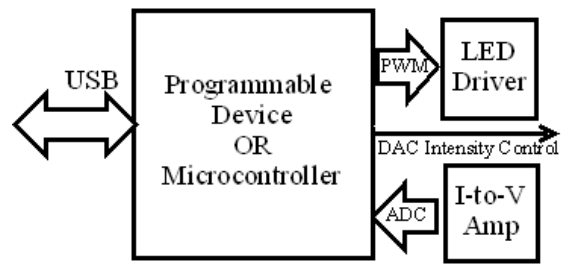

*Fig.1. Basic electronic hardware for Mobile based Pulse Oximeters*

Wavelet transforms are one of the leading techniques used for compression, edge detection, de-noising etc. The paper proposes application of wavelet transform to the pulse-oximeter signals for finding peaks and deriving baseline from the signals to calculate ratio of ratios, the principal factor to find the percentage oxygen saturation.

The following sections discuss mainly the application of wavelet transform to plethysmograph signals viz. infra-red and red signal. Section II discusses in brief about pulse oximeters and its waveforms. It also discusses in brief about wavelet transform. Section III gives a basic understanding of our implementation for finding peaks and baseline from the signals. Section IV discusses about data flow diagrams for the application and suggests future developments.

## II. PRELIMINARY NOTES

This section discusses basics of pulse oximeter and wavelet transforms. We have avoided mathematical equations to discuss the theory, on the contrary have presented practical view that we have understood about both the concepts.

### A. Pulse Oximeters

The standard pulse oximeter [3] works on absorption principal of red and infrared light by the oxygenated and deoxygenated components of the finger. The wavelengths are approximately 660 nm (red) and 910 nm (infrared) produced by LEDs inside the pulse oximeter probe. The photodiode receives the light from the finger and generates current proportional to the absorbed signals. The blood pumped from the heart changes the amount of absorption based on systole and diastole. This produces alternating current (AC) on a fixed bias (DC). The AC and DC of both red and infrared signals are both responsible for calculating the percentage oxygen saturation. The AC signal from infrared (which is stronger than red) is generally considered for calculating the heart rate.

There are lot of challenges involved in instrumentation of probe, adjusting intensity due to motion of the finger, signal conditioning in analog electronics and digital processing. Our application encourages digital signal processing through wavelet transforms and very basic signal conditioning in analog. Secondly the power is also derived from USB source, which is quiet noisy for current to voltage/frequency instrumentation. This might generate noisy raw data and hence the heavy digital signal processing might be required to complement the basic hardware.

*B. Wavelet Transforms*

The wavelet transforms is a very intuitive concept of applying convolution of a signal with a small wave which is shifted from the start of the signal till the end. The small wave, called as wavelet, forms the basis of amplifying a specific frequency based on the scale it uses. As the scale changes, the wavelet transform emphasizes different frequencies in the signal. There are two variants of these, continuous and discrete wavelet transforms. The CWT or continuous wavelet transforms can pin-point the frequency on the time resolution. There are many advantages while using wavelets for signals as compared to Fourier transforms whose resolution is limited to frequency domain.

A mathematical representation of the wavelet transform is given in Figure 2.

$$X_w(a,b) = \frac{1}{\sqrt{a}} \int_{-\infty}^{\infty} x(t)\psi^*\left(\frac{t-b}{a}\right) dt$$

*Fig.2. Continuous wavelet transform*

In the equation the child wavelet $\Psi$ is a version of mother wavelet which is scaled/dilated by $a$ and shifted by $t$. The signal is represented by $x(t)$ and the transform is $X_w$. There are many standard mother wavelets available and one can generate ones own mother wavelet based on the application requirement too. Figure 3 shows the significance of scale in a wavelet called as Gaussian First Order Derivative that we are using in our application.

The wavelet is multiplied with the signal and outputs from the product of signal and shifted versions of the wavelet at a specific scale are added to get wavelet transform. Intuitively, the lower scale will pronounce the higher frequencies in the transform while the higher scale will pronounce the lower frequencies in the transform. This is obviously seen from Figure 3, where same wavelet looks differently when its scale factor is changed. The higher frequency of the signal that can be pronounced is a frequency equal to the sample width.

The Discrete Wavelet Transform works on discretely sampled wavelets. The wavelet is made up of two factors termed as high pass and low pass which generate approximation and details of the signal. The simplest example is of a Haar wavelet. Haar is nothing but a set of {+1, -1}. When the adjacent samples of the signals are averaged, it gives a low pass effect of the signal and their differences are viewed, it gives a high pass effect. Similarly discrete wavelet transform decomposes the signal into levels of details and approximations.

Our application uses both continuous and discrete wavelet transform for (a) finding peaks and (b) finding DC from the pulse oximeter waveforms.

III. IMPLEMETATION WAVELET TRANSFORM

This section briefly describes our work for extracting peaks and baseline from the plethysmograph signal which are crucial for finding percentage oxygen saturation and heart rate.

*A. Heart Rate Detection*

The heart rate from the received infrared waveform is obtained by calculating the peaks of the signal. Tsu-Hsun Fu *et.al.* have de-noised the signal and then performed heart rate calculation on it [4]. This might be good for displaying the clean signal. They use discrete wavelet transforms to remove the noise. We instead use continuous wavelet transform to find the peaks without filtering the signal.

The first order Gaussian derivative wavelet, as the name suggests, is first differentiation of the Gaussian curve. Our application waveform, infrared $SpO_2$ signal, is similar to this. The first order derivative gives a zero at the peak point on the Gaussian curve where the slope tends to zero. Based on this concept, we have found the zero crossings in the wavelet transform to find the peaks (and valleys) of the signal.

Following is the algorithm

1. do wavelet transform of the signal
2. use a threshold which is +/- 30% of the maximum absolute peak from the wavelet transform
3. store the points above the positive threshold as +1 and below negative threshold as -1 in a threshold array, the transitions are visible in the threshold array from -1 to +1.

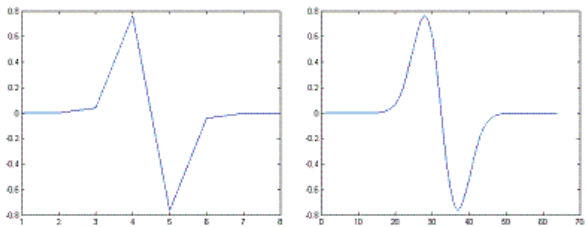

*Fig.3. Scaling of a Gaussian First Order Derivative. On Left the scale factor is 3 and on Right the scale factor is 6*

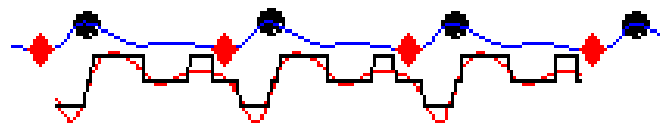

*Fig.4.Wavelet Transform and Peak Detection: The blue signal is the infrared plethysmograph signal with peaks marked as black and valleys marked with red. The black waveform is the threshold array signal and red waveform is the wavelet transform*

4. search for the transition based on threshold array
5. find the zero crossing point
6. if multiple zero crossing points are found, use amplitude comparison from original signal
7. go to 4 until end of the array
8. Store these points into peak array.

TABLE I
SCALE SELECTION AND SNR OF THE SIGNAL

| Scale | Min SNR |
|---|---|
| 4 | 23 |
| 5 | 17 |
| 6 | 10 |
| 7 | 10 |
| 8 | 10 |
| 9 | 08 |

Figure 5 shows a comparison of signal peak-detection at various noise levels. The scale 6 fails at SNR=5 giving the heart rate value of 106 bpm for the signal whose heart rate was 90 bpm. Based on peak valley points the peak height or AC part of the signal can be measured.

### C. Baseline and Motion

Another challenge to execute the raw plethysmograph signal, especially the red signal, is baseline and its variations. The DC signal is important to find the ratio of ratios which is ratio of AC/DC of each signals. This DC may vary slowly over the period of time however it is imperative to find accurate DC signal from the waveforms. The variations in DC are induced by the intensity control circuit and motion. During motion it is considered best to hold the previous SpO2 values and the heart rate. The general DC detection is implemented based on discrete wavelet transform. The general logic is to keep on removing all variations i.e. low pass filtering the signal until only the DC signal remains. The discrete wavelet transforms needs to decompose this signal upto level 8 for getting proper DC of the signal. After the decomposition, only the approximate signal is reconstructed back to find the DC of the original signal.

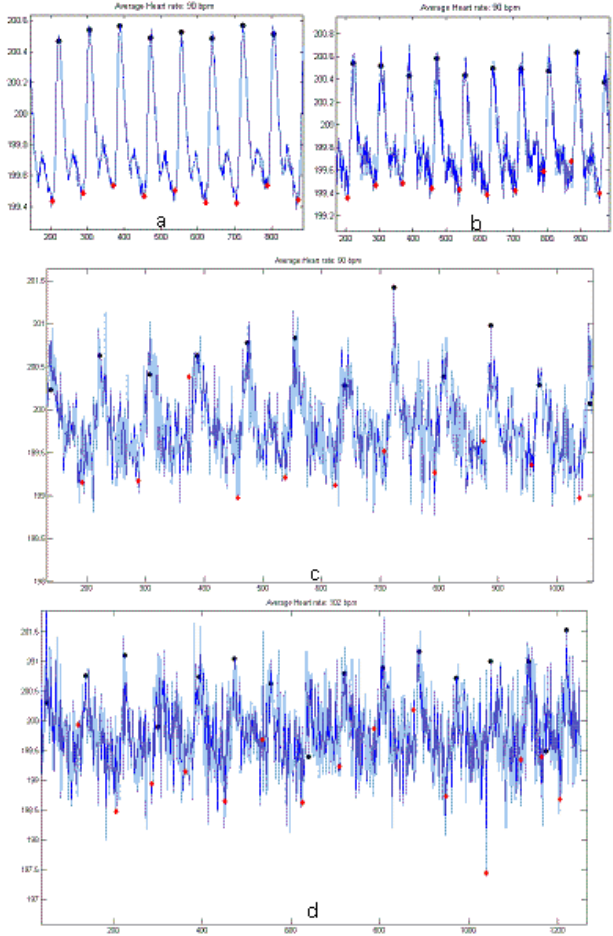

*Fig.5. Application of scale 6 first order Gaussian:
a) SNR=30, b) SNR=20, c) SNR=10, d) SNR=5*

### B. Impact of Noise and Higher Scales

The noise represents high frequency variations in the signal which can be removed by using wavelet transform on higher scales. It was found from the experiments that the scale 6 is the optimum scale which can give accurate heart rate at minimum signal to noise ratio (SNR). As we increase the scale the computational expense of the wavelet transform also increases, hence it is not worth to go beyond scale 6.

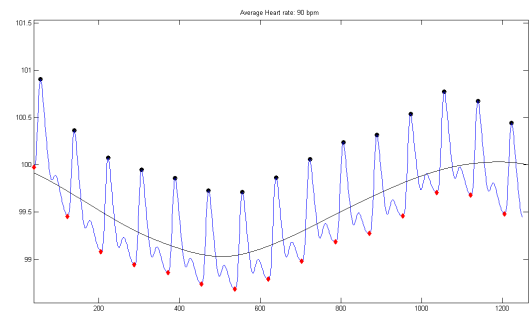

*Fig.6. DC and AC from the plethysmograph waveform*

### D. Noise and Baseline

There is not much challenge in finding AC and DC in signals with noise (upto SNR >= 5) and a DC swing, thanks to inherent properties of wavelet transform. Figure 7 shows both challenges easily been handled by combination of continuous and discrete wavelet transform.

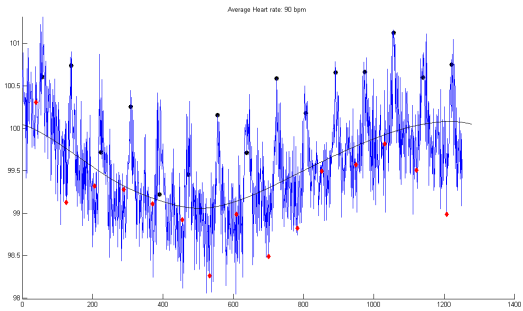

*Fig.7. DC and AC from the noisy plethysmograph waveform*

IV. MOBILE APPLICATION

Now-a-days mobile applications are capable of handling high end graphics and gaming applications. Based on this feature, it might not be a great challenge to perform wavelet transforms on the input signal. The sampling frequency of the pulse-oximeter is also not that great as it might overwhelm the application memory. After the signal is received from the hardware device, based on our algorithms, the mobile application will calculate the percentage oxygen saturation and heart rate of the subject. This can be checked with threshold programmed for individual patients. Heart Rate Variation is also another important critical which can be monitored via this method. The mobile can communicate with the centre mobile device, near a medical staff, using Bluetooth or calling (missed call) on GSM. If no reply is sensed, it can call alarm application with highest volume and a preset ringtone. Figure 8 gives a general draft of the control flow of such an application.

V. CONCLUSION AND FUTURE WORK

Various tools of mobile application programming that are platform independent are being developed. We have ported majority of our code for the application in C# which is yet another platform independent language. The code can be adopted for mobile applications using Windows Mobile and CE operating systems. It can also be ported to Java for gamut of other operating systems such as RIM, Symbian and Andriod. As a part of our on going research we would continue develop such an application as the extension of this work.

ACKNOWLEDGMENT

We wish to acknowledge that a part of our work was done at Maestros Mediline Systems Limited, Navi Mumbai, India; for evaluation of wavelet transforms and would like to thank them to help us introduce to this arena of the wavelet processing.
.

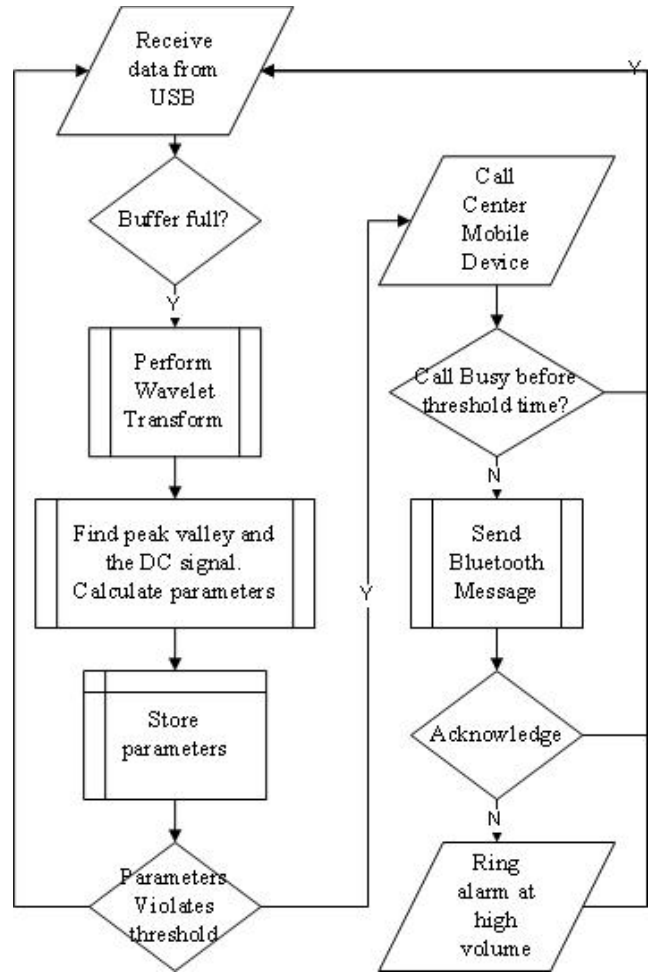

*Fig.8 Flowchart of Mobile Application*